\documentclass{article}

\usepackage[english]{babel}
\usepackage[letterpaper,top=2cm,bottom=2cm,left=3cm,right=3cm,marginparwidth=1.75cm]{geometry}
\usepackage{mathptmx}
\usepackage{sectsty}
\usepackage{amsmath}
\usepackage{amssymb}
\usepackage[numbers]{natbib}
\usepackage{bm}
\usepackage{graphicx} 
\usepackage{hyperref}

\usepackage{subcaption}

\usepackage[format=plain,justification=justified,singlelinecheck=false,font={stretch=1.125,small,sf},labelfont=bf,labelsep=space]{caption}

\usepackage[T1]{fontenc}
\usepackage[version=4]{mhchem}

\usepackage{newtxmath}

\usepackage{gensymb}

\usepackage{gensymb}
\usepackage{comment}
\usepackage{subcaption}

\usepackage{float}

\usepackage{siunitx}             	
\usepackage[version=4]{mhchem}  	
\usepackage{booktabs} 				
\usepackage[flushleft]{threeparttable}	

\title{
	Vacancy-Enhanced \ce{N-N} Bonding and Deep Level Complex Defect Formation in
	\ce{\beta-Ga2O3}
} 

\author{
	Asiyeh Shokri\textit{$^{1,2,}$}$^{\ast}$,
	Yevgen Melikhov\textit{$^{3,}$}$^{\ast}$,
	Yevgen Syryanyy\textit{$^{4}$}, \\
	Maryna Chernyshova\textit{$^{5}$} and 
	Iraida N. Demchenko\textit{$^{6,2}$}
}

\begin{document}

\maketitle

\begin{abstract}
	Nitrogen implantation can make \ce{\beta-Ga2O3} semi-insulating, but the atomic
	origin responsible for carrier trapping remains unresolved. Here, hybrid
	density functional theory is used to examine complexes derived from an
	interstitial-substitutional N pair,	\ce{N_{i9}-N_{OI}}, and effects of nearby O and Ga
	vacancies on \ce{N-N} bonding, stability, and electronic states. Within the
	configurations examined, N atoms favor co-localization. 
	In selected	Ga-vacancy-assisted complexes,
	the \ce{N-N} distance shortens to \qtyrange{1.09}{1.19}{\angstrom}
	and exhibits electron localization consistent with substantial \ce{N-N}
	covalent character. Positive binding energies indicate favorable
	association relative to separated defects, although many of these
	complexes have high formation energies and are therefore more likely to
	be metastable products of implantation than abundant equilibrium
	defects. Calculated N~$K$-edge XANES spectra for two selected
	Ga-defect-assisted \ce{N-N} complexes reveal a pronounced sensitivity of the
	near-edge $\pi^*$ resonance to the local defect environment, providing an
	atomistic perspective on the experimentally observed evolution toward
	molecular \ce{N_{2}} during annealing. Projected densities of states and
	spin-density distributions show localized N-$2p$- and O-$2p$-derived states
	well inside the band gap. These vacancy-assisted N complexes therefore
	provide an atomistic mechanism for deep carrier trapping and current
	blocking rather than shallow acceptor doping in \ce{\beta-Ga2O3}.
\end{abstract}

\footnotetext{\textit{$^{1}$}~International Research Centre Magtop, 
	Institute of Physics 
	Polish Academy of Sciences, Aleja~Lotnik\'ow 32/46, 02-668
	Warsaw, Poland}

\footnotetext{\textit{$^{2}$}~Institute of Plasma Physics and Laser Microfusion,
	ul.~Hery 23, 01-497 Warsaw, Poland}

\footnotetext{\textit{$^{3}$}~Institute of Fundamental Technological Research
	Polish Academy of Sciences, ul.~Pawinskiego 5b, 02-106 Warsaw, Poland}

\footnotetext{\textit{$^{4}$}~Institute of Microelectronics and Optoelectronics,
	Warsaw University of Technology, ul.~Koszykowa 75, 00-662 Warsaw, Poland
}

\footnotetext{\textit{$^{5}$}~National Center for Nuclear Research, Andrzeja
	Sołtana 7, 05-400 Otwock, Poland
}

\footnotetext{\textit{$^{6}$}~The Centre for Advanced Materials and Technologies,
	CEZAMAT at the Warsaw University of Technology, 19 Poleczki St, Warsaw 02-822, Poland
}

\footnotetext{$^{\ast}$~corresponding authors: 
		Y.M. \url{Yevgen.Melikhov@ippt.pan.pl}, 
		A.S. \url{AShokri@magtop.ifpan.edu.pl}, \url{Asiye.Shokri@gmail.com}
}

\vspace{0.5cm}

\textbf{Highlights}

\begin{itemize}
	\item
	Nitrogen atoms exhibit a strong tendency toward co-localization in
	\ce{\beta-Ga2O3}, promoting \ce{N-N} interaction.
	\item
	Vacancy-assisted configurations enhance \ce{N-N} bonding and favor the
	formation of complex defect structures near Ga vacancies.
	\item
	N-related defect complexes introduce deep electronic states that act
	as carrier traps, leading to charge blocking rather than shallow
	$p$-type conductivity.
\end{itemize}

	\ce{N_{2}} in \ce{Ga2O3};
	XANES
	
\section{Introduction}
\label{section1}

	\ce{\beta-Ga2O3} is an ultrawide-band-gap semiconductor of interest for
	high-voltage and high-power electronic devices. Nitrogen incorporation
	has been investigated as a means of controlling its electronic
	properties, particularly because N substituting for O is nominally
	acceptor-like \cite{1,2,3,4}. Both theoretical and experimental
	studies, however, indicate that N-related acceptor levels are generally
	deep within the band gap. Consequently, N incorporation tends to
	compensate carriers rather than produce mobile
	holes \cite{5,6}.

	Nitrogen introduced under nonequilibrium conditions can alter the
	electrical behavior more strongly than an isolated substitutional defect
	would suggest \cite{7}. In particular, N implantation
	increases the resistivity of \ce{\beta-Ga2O3} and has been used to form
	semi-insulating current-blocking regions in vertical
	devices \cite{8,9}.  However, the microscopic defects
	responsible for this behavior have not yet been identified conclusively.

	The atomic configuration of N is expected to depend strongly on the
	incorporation route. While Near-equilibrium growth may favor isolated
	defects, the high-energy collision cascades of ion implantation generate
	high concentrations of vacancies and interstitials that can become
	kinetically trapped or combine into metastable
	complexes \cite{10,11}. Interactions among multiple N atoms
	and native defects can therefore modify both the local bonding and the
	electronic structure \cite{2,7}. Spectroscopic evidence for
	molecular \ce{N_{2}} in N-implanted \ce{\beta-Ga2O3} has further motivated consideration
	of \ce{N-N}-containing configurations \cite{7,12}. In particular,
	temperature-dependent N~$K$-edge XANES measurements have revealed the
	progressive formation of molecular \ce{N_{2}} upon annealing of N-implanted
	\ce{\beta-Ga2O3} \cite{7}. However, the atomic configurations of
	nitrogen in the highly damaged as-implanted material, and the possible
	role of implantation-induced intrinsic defects in the subsequent
	evolution of \ce{N-N} bonding, remain insufficiently understood. Related \ce{N-N}
	complexes have also been discussed in \ce{ZnO} \cite{13,14}.
	
	Our previous hybrid-functional calculations identified
	\ce{N_{i9}-N_{OI}}, comprising an N interstitial at
	the \ce{i9} site and an N atom substituting at the \ce{O_{I}} site,
	as the lowest-energy arrangement among the N interstitial-substitutional
	configurations examined under Ga-rich conditions \cite{15}.
	Its \ce{N-N} distance of approximately \qty{1.27}{\angstrom} is longer than that of an
	isolated \ce{N_{2}} molecule, and the electron localization function (ELF)
	revealed appreciable interaction between one N atom and neighboring Ga
	atoms. The complex therefore has only partial molecular
	character \cite{15}.
	
	We hypothesize that nearby native vacancies and associated atomic
	rearrangements may modify the local coordination and electronic
	environment of incorporated nitrogen, thereby influencing \ce{N-N} bonding,
	defect stability, and the formation of localized electronic states.
	While Background hydrogen can also influence defect chemistry in oxides
	through passivation and complex formation \cite{16}, the
	present work is deliberately restricted to the intrinsic interaction
	between N and native O or Ga vacancies to isolate their fundamental
	effects.
	
	Using density functional theory and XANES simulations, our objectives
	are to determine: (i) whether vacancy-assisted relaxation strengthens
	\ce{N-N} bonding, (ii) whether the resulting complexes are energetically
	favorable relative to their isolated constituents, (iii) whether their
	electronic states can account for carrier trapping in implanted \ce{\beta-Ga2O3},
	and (iv) how the local Ga-defect environment influences the N~$K$-edge
	XANES signatures of selected \ce{N-N} complexes in comparison with
	experimental observations.

\section{Methods}
\label{section2}

\subsection{Computational Methods}
\label{section2.1}

	Spin-polarized first-principles calculations were performed with the
	projector-augmented-wave method implemented in VASP~6.3.2 \cite{17, 18, 19, 20}.
	A plane-wave cutoff energy of \qty{510}{\eV} was
	used. Initial calculations employed the Perdew--Burke--Ernzerhof
	generalized-gradient approximation \cite{21}.
	Defect formation energies, relaxed structures, and electronic properties were evaluated
	using a modified Heyd--Scuseria--Ernzerhof screened hybrid functional with
	a screening parameter of \qty{0.2}{\angstrom^{-1}} and an exact-exchange fraction of
	\qty{0.35}{} \cite{22}. 
	This parametrization gives a band gap of 
	\qty{4.91}{\eV} for pristine \ce{\beta-Ga2O3} and lattice parameters consistent with
	experiment.

	Defects were modeled in a \qtyproduct{1 x 4 x 2}{} conventional-cell supercell containing 160 atoms. 
	The fixed supercell parameters were 
	$a = \qty{12.22}{\angstrom}$, 
	$b = \qty{12.112}{\angstrom}$, 
	$c = \qty{11.58}{\angstrom}$, and 
	$\beta=\qty{103.684}{\degree}$.
	Brillouin-zone integrations used $\Gamma$-centered \qtyproduct{2 x 8 x 4}{} and \qtyproduct{1 x 1 x 1}{}
	k-point meshes for the 20-atom conventional cell and the 160-atom
	supercell, respectively. Gaussian smearing with a width of \qty{0.05}{\eV} was
	applied. Atomic positions were relaxed at fixed lattice parameters until
	all residual forces were below \qty{0.03}{\eV.\angstrom^{-1}}. The electronic
	self-consistency threshold was \qty{1e-4}{\eV}. For each defect, the spin
	configuration was allowed to relax self-consistently, and the
	lowest-energy magnetic solution obtained was used in the analysis.

	The formation energy of defect \ce{X} in charge state $q$ was calculated as \cite{23}
	\begin{equation}
		E_{f}\left( X^{q} \right) = 
		\ E_{tot}\left( X^{q} \right) - 
		E_{tot}(bulk) - 
		\sum_{i}^{}{n_{i}\mu_{i}} + 
		q\ \left( E_{F} + E_{VBM} \right) + \ E^{q}_{corr},
	\end{equation}	
	where $E_{tot}\left( X^{q} \right)$ and $E_{tot}(bulk)$ are the
	total energies of the defective and pristine supercells, respectively.
	The quantity $n_{i}$ is the number of atoms of species $i$ added to
	($n_{i} > 0$) or removed from ($n_{i} < 0$) the supercell, and
	$\mu_{i}$ is its chemical potential. 
	The Fermi energy $E_{F}$ is referenced to the valence-band maximum $E_{VBM}$. 
	The correction $E_{corr}^{q}$ was evaluated using the Freysoldt--Neugebauer--Van de
	Walle scheme with a dielectric constant of \qty{10}{} to account for spurious
	electrostatic interactions and potential alignment in charged supercells \cite{23}.
	Charged cells were neutralized by a uniform compensating background.

	The elemental chemical potentials were written as
	$\mu_{i} = \mu_{i}^{0} + \Delta\mu_{i}$, where $\mu_{i}^{0}$ is the
	energy of the elemental reference and $\Delta\mu_{i} \leq 0$.
	Stability of \ce{\beta-Ga2O3} requires
	\begin{equation}
		2\Delta\mu_{\text{Ga}} + 3\Delta\mu_{\text{O}} = 
		\Delta H_{f}\left( \text{Ga}_{2}\text{O}_{3} \right) .
	\end{equation}

	The Ga-rich limit was defined by $\Delta\mu_{\text{Ga}} = 0$, whereas the
	O-rich limit was defined by $\Delta\mu_{\text{O}} = 0$. The N reference was
	one-half of the total energy of an isolated \ce{N_{2}} molecule. The accessible
	N chemical potential was additionally constrained to prevent \ce{GaN}
	precipitation,
	\begin{equation}
		\Delta\mu_{\text{Ga}} + \Delta\mu_{\text{N}} \leq \Delta H_{f}(\text{GaN}).
	\end{equation}
	Thermodynamic transition levels were obtained from \cite{24}
	\begin{equation}
		\varepsilon(q_{1}/q_{2}) = \frac{E_{f}( X^{q_{1}}; E_{F} = 0) - E_{f}(X^{q_{2}}; E_{F} = 0)}{q_{2} - q_{1}}.
	\end{equation}
	\ce{\beta-Ga2O3} contains two crystallographically distinct 
	Ga sites (\ce{Ga_{I}} and \ce{Ga_{II}}) and 
	three O sites (\ce{O_{I}}, \ce{O_{II}}, and \ce{O_{III}}). 
	A vacancy is denoted by \ce{V} with the corresponding site as a subscript 
	(for example, \ce{V_{OI}} or \ce{V_{GaII}}), and substitutional \ce{N} at \ce{O_{I}} is denoted \ce{N_{OI}}.
	The N interstitial at the i9 cavity is denoted \ce{N_{i9}}.
	\ce{{FP}_{GaI}} denotes the \ce{Ga_{I}} Frenkel pair-like rearrangement 
	produced when a \ce{Ga_{I}} atom leaves its lattice site and moves to a nearby interstitial position.

\subsection{Experimental Methods}
\label{section2.2}

	X-ray absorption experiments were performed at the SOLARIS synchrotron
	using the PIRX beamline. All in situ/ex situ X-ray absorption near-edge
	structure (XANES) measurements were carried out at room temperature (RT)
	on samples subjected to stepwise annealing over the temperature range of
	\qtyrange{320}{1100}{\celsius}. For the purposes of the present study, the analysis is
	restricted to the spectra corresponding to two representative states:
	the as-implanted sample measured at RT and the sample subsequently
	annealed at \qty{725}{\celsius}.

	The XANES spectra were acquired in total fluorescence yield (TFY) mode
	by recording the fluorescence signal from the samples while scanning the
	photon energy across the N~$K$-edge region. 
	Annealing was performed under ultrahigh-vacuum conditions 
	(pressure below \qty{6e-9}{mbar}) in a dedicated chamber. 
	Depending on the heating mode, the heating rate varied between 
	\qty{3}{} and \qty{10}{\celsius.s^{-1}} using either radiative heating alone or
	a combination of radiative and electron-beam heating. 
	At each temperature step, the samples were annealed for approximately \qty{10}{min}.
	Below \qty{400}{\celsius}, the sample temperature was monitored using a calibrated
	thermocouple mounted on the manipulator, whereas above \qty{400}{\celsius} it was
	monitored using an Optris CSvision R2ML pyrometer.
	
	After normalization to the incident photon flux, the recorded N~$K$-edge
	XANES spectra were subjected to subtraction of a linear background
	fitted to the flat pre-edge region. For quantitative comparison, the
	spectra were subsequently normalized to the atomic limit in the region
	approximately \qty{30}{eV} above the absorption edge.

\section{Results and Discussion}
\label{section3}

\subsection{Nitrogen Co-localization and Vacancy-Assisted \ce{N-N} Bonding}
\label{section3.1}

\begin{figure}[h!]
		\centering
		\begin{subfigure}{0.5\textwidth}
			\centering
			\includegraphics[width=\textwidth]{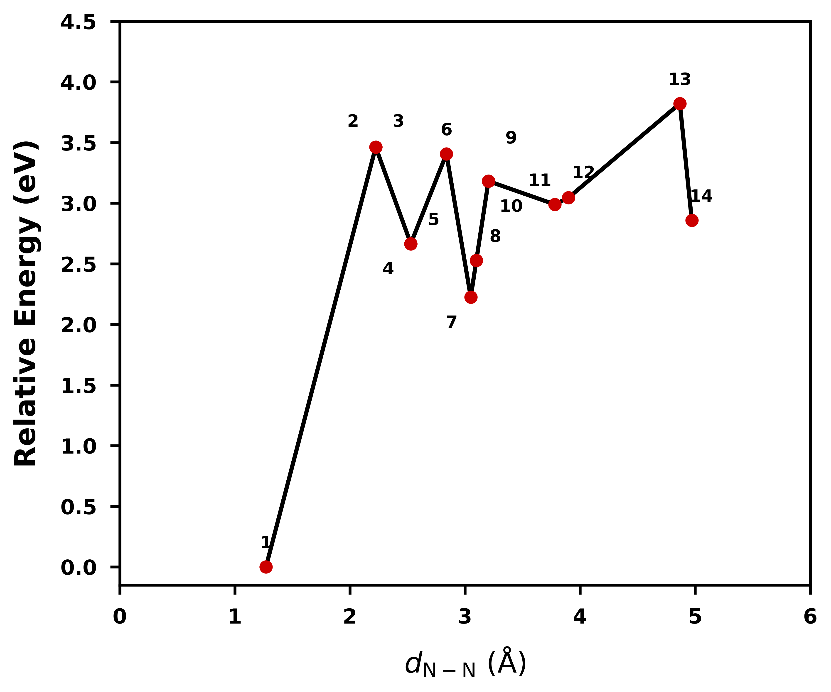}
			\caption{}
			\label{figure1a}
		\end{subfigure}
		\hspace{0.3cm}
		\begin{subfigure}{0.4\textwidth}
			\centering
			\includegraphics[width=\textwidth]{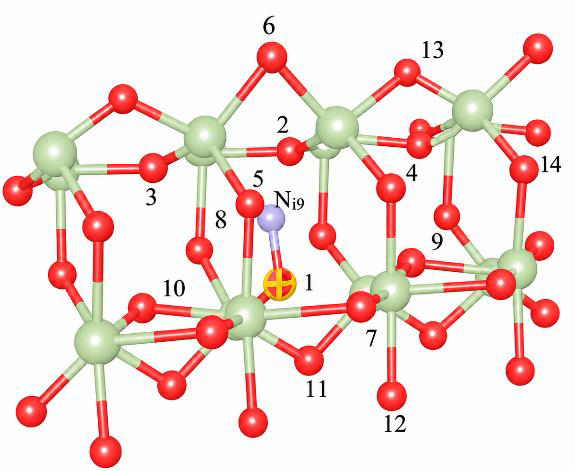}
			\caption{}
			\label{figure1b}
		\end{subfigure}
		\caption{
(a) Relative energies, $\Delta E$, of the
\ce{N_{i9}-N_{Ox}} configurations as a function of
the \ce{N-N} separation, $d_{\ce{N-N}}$. 
Labels 1-14 identify the corresponding O sites shown in (b). 
Grouped labels (2,3), (4,5), and (9,10) indicate configurations with 
nearly identical \ce{N-N} separations and
relative energies, whose markers overlap at the scale shown. 
The \ce{N_{i9}-N_{O1}} configuration defines the
zero-energy reference. 
(b) Local atomic environment around
\ce{N_{i9}}, showing the O sites considered for substitution by N. 
The reference \ce{O_{1}} site is indicated by a yellow cross.
	} 
		\label{figure1}
	\end{figure}

	Starting from the previously identified
	\ce{N_{i9}-N_{OI}} complex \cite{15}, 
	substitutional N atoms were placed at several O sites surrounding the
	\ce{N_{i9}} interstitial. 
	The structures were then fully relaxed
	to compare their \ce{N-N} separations and total energies. 
	Within this set of
	configurations, the shortest \ce{N-N} separation coincides with the lowest
	relative energy (Figure~\ref{figure1}).
	\ce{N_{i9}-N_{OI}} is therefore the ground-state arrangement among the
	\ce{N_{i9}-N_{Ox}} configurations examined. 
	At larger separations, the relative energy is nonmonotonic, showing that
	the local coordination environment and lattice relaxation contribute in
	addition to the direct \ce{N-N} interaction.

	Although these calculations show that nitrogen atoms implanted in
	\ce{Ga_{2}O_{3}} prefer to cluster together, the
	resulting defect complex cannot be characterized as a true nitrogen
	molecule based on its bond length and electron localization function
	(ELF) analysis \cite{15}.
	Consequently, this configuration
	differs from the \ce{N_{2}} molecular species reported in
	experimental studies of N-implanted \ce{Ga_{2}O_{3}} \cite{7, 12}.
	To determine whether native defects strengthen the \ce{N-N} interaction,
	\ce{V_{OI}}, \ce{V_{GaI}}, \ce{V_{GaII}}, and
	the \ce{V_{GaI}-V_{GaII}} pair were introduced near
	\ce{N_{i9}-N_{OI}}.
	
	All defect models were structurally relaxed in each charge state
	considered in the formation energy calculations. Because charging may
	also modify the local \ce{N-N} interaction, the \ce{N-N} distances were extracted
	from the independently relaxed structures used in the formation energy
	calculations and are summarized in Table~\ref{table1}.

\begin{table}[htbp]
\centering
\begin{threeparttable}
\caption{\ce{N-N} Distances Calculated with the Modified HSE Functional}
\label{table1}
\begin{tabular}{l|c|c|c|c|c}
	\toprule
	\textbf{Defect configuration} & 	\textbf{$q = -2$} & 
										\textbf{$q = -1$} & 
										\textbf{$q = 0$} & 
										\textbf{$q = +1$} & 
										\textbf{$q = +2$} \\
	\midrule
	\ce{N_{i9}-N_{OI}} 						& 1.269 & 1.268 & 1.270 & 1.174 & 1.215 \\
	\ce{N_{i9}-N_{OI}-V_{OI}}				& 1.456 & 1.363 & 1.340 & 1.348 & 1.254 \\
	\ce{N_{i9}-N_{OI}-V_{GaI}}				& - & 1.110 & 1.190 & 1.094 & 1.103 \\
	\ce{N_{i9}-N_{OI}-V_{GaII}}				& 1.240 & Transformed to \ce{{FP}_{GaI}} \tnote{a} & 1.270 & 1.269 & 1.269 \\
	\ce{N_{i9}-N_{OI}-V_{GaII}-{FP}_{GaI}}	& - & 1.094 \tnote{a} & 1.090 & 1.090 & - \\
    	\ce{N_{i9}-N_{OI}-V_{GaI}-V_{GaII}}	& - & 1.085 & 1.100 & 1.090 & - \\
	Isolated \ce{N_{2}}, DFT				& - & - & 1.097 & - & - \\
    	Isolated \ce{N_{2}}, experiment \tnote{b}	& - & - & 1.098 & - & - \\
	\bottomrule
\end{tabular}
\begin{tablenotes}
	\small
	\item[a] 
		During relaxation at $q = -1$, the \ce{N_{i9}-N_{OI}-V_{GaII}} 
		precursor underwent a qualitative structural rearrangement 
		in which a neighboring \ce{Ga_{I}} atom moved to an interstitial 
		position, producing \ce{N_{i9}-N_{OI}-V_{GaII}-{FP}_{GaI}}. 
		Therefore, no bond length is assigned to an unreconstructed 
		\ce{N_{i9}-N_{OI}-V_{GaII}} structure at this charge state.
	\item[b] 
		Ref.~\cite{Huber1979}.
\end{tablenotes}
\end{threeparttable}
\end{table}

	The \ce{N-N} separation exhibits a clear but configuration-dependent response to charge. 
	In \ce{N_{i9}-N_{OI}}, the distance
	remains close to \qty{1.27}{\angstrom} from $q= -2$ to $q= 0$ and decreases in the positive
	charge states. The largest variation is found for
	\ce{N_{i9}-N_{OI}-V_{OI}}, for which
	the \ce{N-N} distance ranges from \qty{1.456}{\angstrom} at $q= -2$ to \qty{1.254}{\angstrom} at $q= +2$. 
	By contrast, the \ce{{FP}_{GaI}}- and double-Ga-vacancy-containing
	configurations retain short \ce{N-N} distances of approximately \qtyrange{1.085}{1.100}{\angstrom}
	throughout the charge states examined, comparable to the calculated and
	experimental distances of isolated \ce{N_2}. Although charging changes the \ce{N-N}
	separation in several configurations, only
	\ce{N_{i9}-N_{OI}-V_{GaII}} at $q= -1$
	undergoes a qualitative reconstruction of its surrounding atomic environment. 
	This reconstruction generates the additional
	\ce{N_{i9}-N_{OI}-V_{GaII}-{FP}_{GaI}} model included in the subsequent analysis.

\begin{figure}[h!]
		\centering
		\begin{subfigure}{0.45\textwidth}
			\centering
			\includegraphics[width=\textwidth]{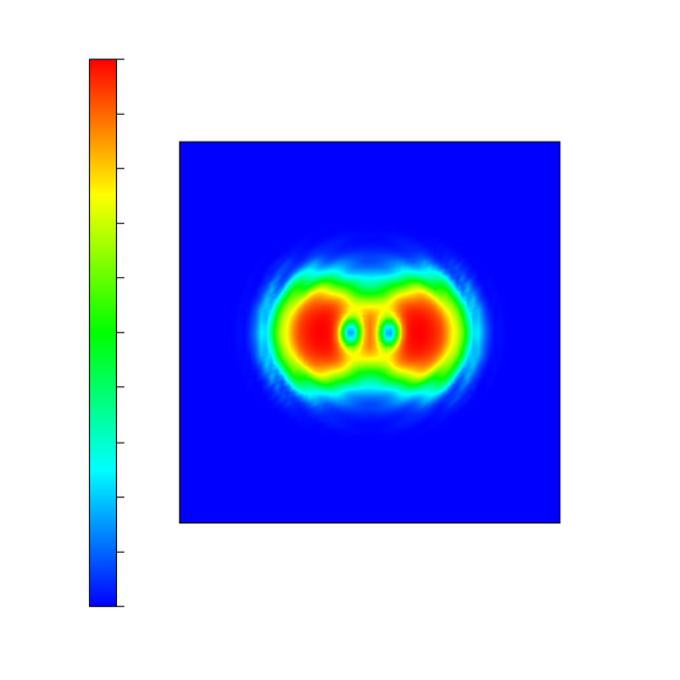}
			\caption{}
			\label{figure2a}
		\end{subfigure}
		\hspace{0.3cm}
		\begin{subfigure}{0.45\textwidth}
			\centering
			\includegraphics[width=\textwidth]{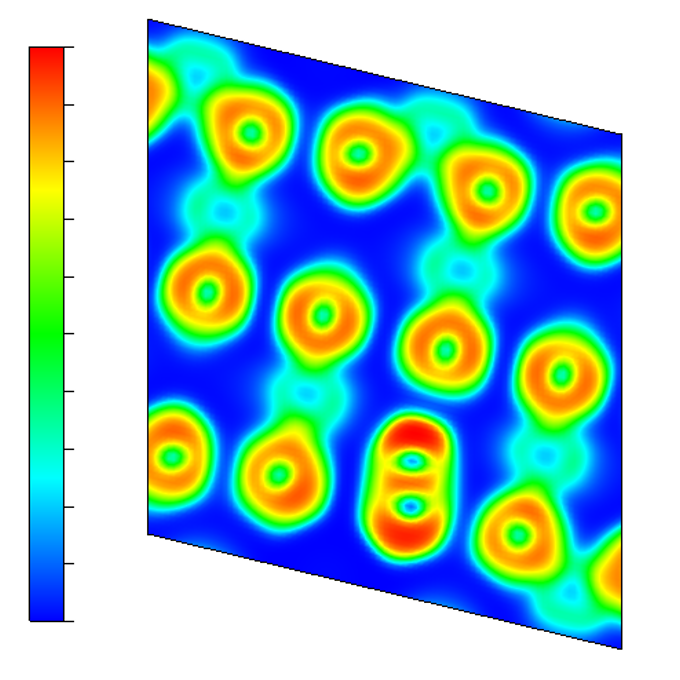}
			\caption{}
			\label{figure2b}
		\end{subfigure}
		\caption{
			Electron localization function for:
			(a) an isolated \ce{N_{2}} molecule and 
			(b) \ce{N_{i9}-N_{OI}-V_{GaI}}. 
			The same color scale is used in both panels.
		} 
		\label{figure2}
	\end{figure}

	The ELF provides complementary information. 
	Figure~\ref{figure2} shows pronounced
	electron localization between the two N atoms in
	\ce{N_{i9}-N_{OI}-V_{GaI}}, similar to
	the localization between the atoms of isolated \ce{N_{2}}.
	This result supports
	substantial \ce{N-N} covalent character in the embedded complex, while the
	remaining interaction with the \ce{\beta-Ga2O3} lattice distinguishes it from a
	free \ce{N_{2}} molecule.

	The structural results presented above show that N atoms tend to
	co-localize near the \ce{N_{i9}} site, with this tendency
	becoming more pronounced in the presence of intrinsic vacancies. In
	selected vacancy-assisted configurations, the shorter \ce{N-N} distances and
	increased electron localization between the N atoms indicate stronger
	\ce{N-N} bonding. These structural changes may influence both the stability
	and electronic properties of the complexes. The following section
	therefore examines the formation and binding energies of these
	configurations, while their defect-induced electronic states are
	discussed subsequently.

\subsection{Defect Stability and Spectroscopic Signatures}
\label{section3.2}

\begin{figure}[h!]
		\centering
		\includegraphics[width=0.8\textwidth]{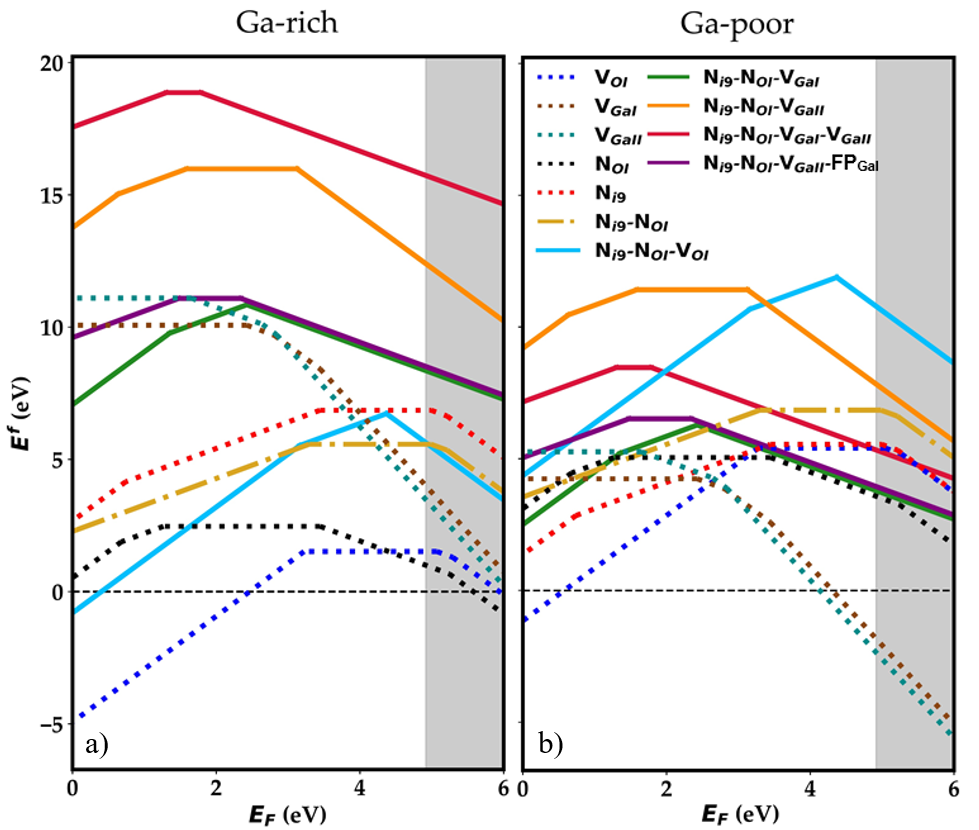}
		\caption{
Formation energies of isolated defects and N-related
complexes in \ce{\beta-Ga2O3} under (a) Ga-rich and (b) Ga-poor conditions. 
The slopes indicate defect charge states, and the vertical intersections
identify thermodynamic transition levels.
		} 
		\label{figure3}
\end{figure}

	Formation energies of the selected defect configurations were calculated
	for charge states from -2 to +2 (Figure~\ref{figure3}). Their dependence on Fermi
	energy confirms that the preferred charge state and the relative
	stability of each complex depend on the growth limit and electronic
	chemical potential. Ga-vacancy-containing complexes become relatively
	more favorable under Ga-poor conditions, whereas the
	O-vacancy-containing complex is less favorable. Relative to isolated
	\ce{N_{OI}}, the calculated $\epsilon(0/-)$ levels of
	several complexes shift toward the valence-band maximum, corresponding
	to a smaller acceptor ionization energy.
	\ce{N_{i9}-N_{OI}-V_{OI}} is an
	exception: its transition level lies deeper in the gap because the
	donor-like O vacancy supplies electrons and changes the local bonding.
	None of the complexes produces a shallow acceptor level in the present
	calculations.

	Importantly, the configurations with the shortest \ce{N-N} distances do not
	necessarily have the lowest formation energies. Thus, enhanced \ce{N-N}
	bonding and thermodynamic favorability do not follow the same trend.
	Furthermore, as discussed in the previous section, in the -1 charge
	state, \ce{N_{i9}-N_{OI}-V_{GaII}} relaxes to \ce{N_{i9}-N_{OI}-V_{GaII}-{FP}_{GaI}}. 
	The formation energy of the rearranged
	structure is similar to that of
	\ce{N_{i9}-N_{OI}-V_{GaI}} in the same
	charge state, but its \ce{N-N} distance is closer to that of isolated \ce{N_{2}}.
	Unlike \ce{N_{i9}-N_{OI}-V_{GaI}} configuration,
	\ce{N_{i9}-N_{OI}-V_{GaII}-{FP}_{GaI}}
	is also locally stable in the neutral charge state.

	The binding energy was evaluated only for neutral defect configurations
	and was used to compare each complex with the same constituent defects
	calculated separately. For a neutral complex $C$ formed from
	corresponding isolated neutral defects $D_{j}$, the binding energy is
	defined as \cite{25,26}
	\begin{equation}
		\label{equation5}
		E_{B}(C) = \sum_{j}^{}E_{f}\left( D_{j} \right) - E_{f}(C),
	\end{equation}
	A positive $E_{B}$ indicates that the formation of the complex is
	thermodynamically stable relative to the corresponding isolated defects.
	
	All complexes in Table~\ref{table2} have positive neutral-state binding energies.
	The values are identical at the Ga-rich and Ga-poor limits because the
	chemical-potential terms cancel out when the isolated constituents and
	the complex have the same overall stoichiometry. The large binding
	energies of the vacancy-assisted structures indicate strong short-range
	chemical interactions and relaxation after the defects are brought into
	proximity. They do not imply high equilibrium concentrations: several
	complexes also have high absolute formation energies and are therefore
	better interpreted as bound, metastable structures accessible within the
	heavily damaged local structure produced during ion implantation.

\begin{table}[htbp]
\centering
\caption{Neutral-State Formation and Binding Energies of Isolated Defects and N-Related Complexes}
\label{table2}
\begin{tabular}{l|c|c|c}
	\toprule
	\textbf{Model} & \textbf{$E_{\text{f}}$, Ga-rich (eV)} & 
	\textbf{$E_{\text{f}}$, Ga-poor (eV)} & \textbf{$E_{\text{B}}$ (eV)} \\
	\midrule
	\ce{N_{OI}}									& 2.468 & 5.046 & - \\
	\ce{N_{i9}}									& 6.856 & 5.546 & - \\
	\ce{V_{GaI}}								& 10.071 & 4.238 & - \\						
	\ce{V_{GaII}}								& 11.091 & 5.258 & - \\
	\ce{V_{OI}}									& 1.516 & 5.404 & - \\
	\ce{{FP}_{GaI}} 								& 1.605 & 1.605 & - \\
	\ce{N_{i9}-N_{OI}}							& 5.572 & 6.840 & 3.752 \\
	\ce{N_{i9}-N_{OI}-V_{OI}}					& 6.799 & 11.955 & 4.041 \\
	\ce{N_{i9}-N_{OI}-V_{GaI}}					& 13.084 & 8.519 & 6.311 \\
	\ce{N_{i9}-N_{OI}-V_{GaII}}					& 15.978 & 11.413 & 4.437 \\
	\ce{N_{i9}-N_{OI}-V_{GaII}-{FP}_{GaI}}		& 11.081 & 6.516 & 10.939 \\
    	\ce{N_{i9}-N_{OI}-V_{GaI}-V_{GaII}}		& 18.859 & 8.461 & 11.627 \\
	\bottomrule
\end{tabular}
\end{table}

	All values correspond to neutral defects; they are therefore independent
	of $E_{F}$. Binding energies were evaluated using the isolated
	constituents listed in Table~\ref{table2} and equation~\eqref{equation5}.
	
	The formation-energy analysis describes the thermodynamic favorability
	of the defect complexes, while the binding energies indicate whether the
	neutral defects preferentially associate rather than remain isolated.
	These quantities, however, do not reveal the energetic position, orbital
	character, or spatial localization of the defect-induced states.

\begin{figure}[h!]
		\centering
		\includegraphics[width=0.5\textwidth]{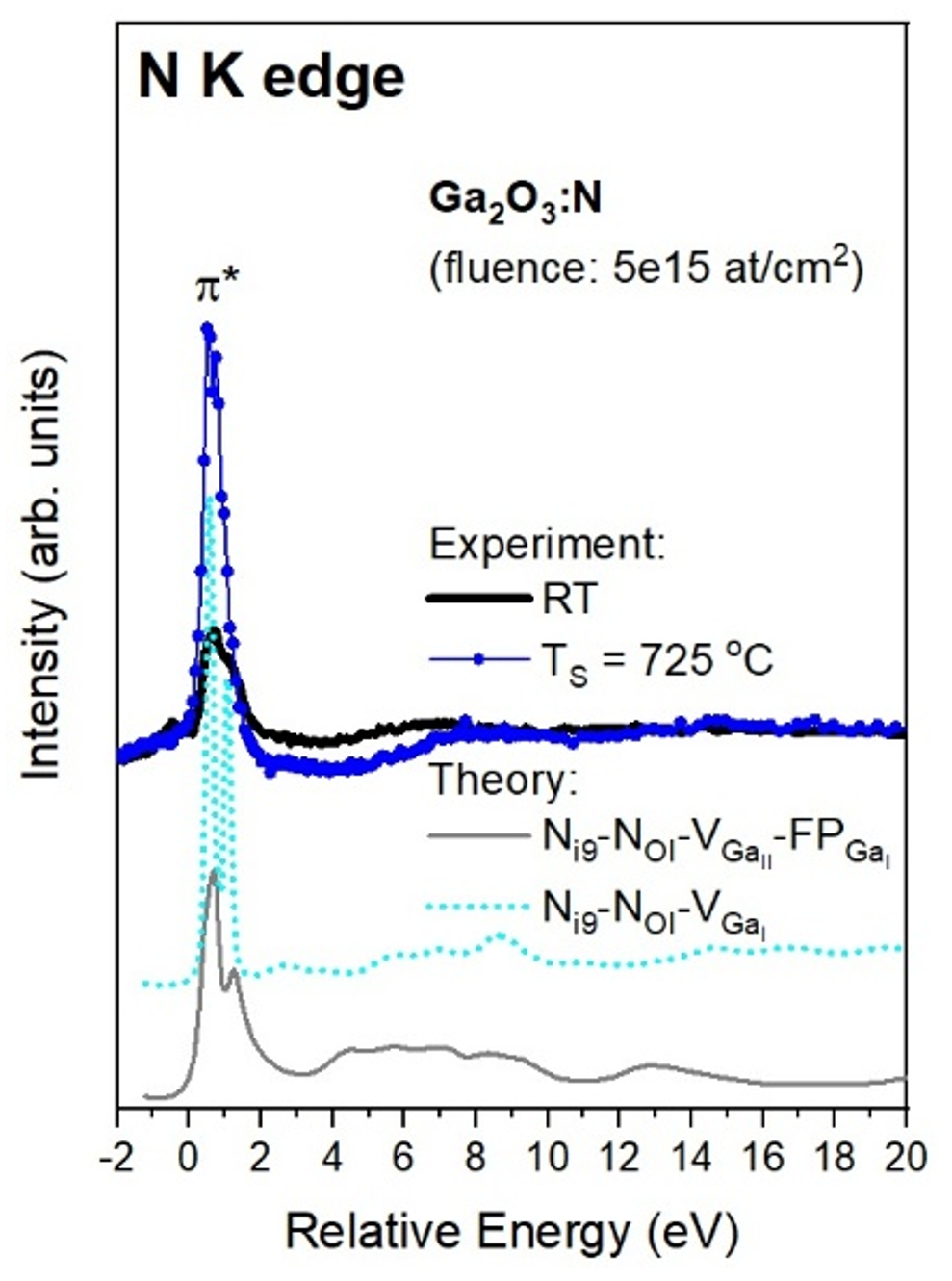}
		\caption{
Spectroscopic signatures of Ga-defect-assisted \ce{N-N}
complexes in \ce{\beta-Ga2O3}. 
Experimental N~$K$-edge XANES spectra of N-implanted
\ce{\beta-Ga2O3} measured in the as-implanted state (RT) and after annealing at
$T_{S}=\qty{725}{\celsius}$  are compared with calculated spectra for two configurations
derived from the \ce{N_{i9}-N_{OI}} complex: 
one containing a Ga vacancy at the \ce{Ga_{I}} site
(\ce{N_{i9}-N_{OI}-V_{GaI}}) and the
other containing a \ce{Ga_{II}} vacancy together with a Ga
Frenkel-like rearrangement
(\ce{N_{i9}-N_{OI}-V_{GaII}-FP_{GaI}}).
The pronounced near-edge $\pi^*$ resonance is associated with \ce{N-N} bonding,
while its calculated intensity and line shape depend on the local
Ga-defect environment. 
Experimental spectra are adapted from Ref.~\cite{7}.
}
		\label{figure4}
\end{figure}

	To further assess whether the vacancy-assisted \ce{N-N} configurations
	considered here are compatible with the local nitrogen environment
	produced by ion implantation, N~$K$-edge XANES spectra were calculated for
	two selected Ga-vacancy-assisted complexes:
	\ce{N_{i9}-N_{OI}-V_{GaI}} and
	\ce{N_{i9}-N_{OI}-V_{GaII}-{FP}_{GaI}}
	(Figure~\ref{figure4}). These configurations were selected based on their relatively
	low formation energies among the Ga-defect-containing complexes over the
	relevant range of Fermi energies under Ga-poor conditions (Figure~\hyperref[figure3]{3b}).
	The latter configuration originates from the structural rearrangement of
	\ce{N_{i9}-N_{OI}-V_{GaII}}
	during relaxation in the $q = -1$ charge state, when a neighboring
	\ce{Ga_{I}} atom moves into an interstitial position, leaving an
	additional Ga vacancy. Thus, this complex contains both a
	\ce{V_{GaII}} and a Ga Frenkel-like rearrangement. The N~$K$-edge
	XANES spectra were calculated from the relaxed atomic structures using
	the FDMNES code \cite{27}.
	The formation of Ga Frenkel pairs
	under displacement damage in \ce{\beta-Ga2O3} is supported by recent atomistic
	and first-principles studies \cite{28,29,30}.
	 Both selected
	complexes exhibit positive binding energies and short \ce{N-N} distances,
	supporting their consideration as possible metastable configurations
	formed during ion implantation. Importantly, both are derived from the
	\ce{N_{i9}-N_{OI}} motif and therefore differ from
	the purely interstitial \ce{N_{2}} configuration considered
	previously \cite{7, 15}.

	The microscopic state of the implantation-modified \ce{\beta-Ga2O3} matrix is,
	however, expected to be considerably more complex than any individual
	defect model considered here. Ion implantation drives the material far
	from thermodynamic equilibrium and generates a heterogeneous local
	environment comprising implantation-induced vacancies and interstitials,
	Frenkel pairs, defect complexes, and varying degrees of local structural
	disorder \cite{28,29,30}. The implanted nitrogen atoms can
	therefore experience a range of distinct local atomic environments, and
	several locally stable or metastable N-containing configurations may
	coexist and contribute to the measured N~$K$-edge response. Accordingly,
	the two configurations examined here should not be regarded as unique
	structural assignments of the implanted layer, but rather as
	representative local environments whose calculated spectra provide
	characteristic signatures of \ce{N-N} bonding influenced by the surrounding
	Ga defects.

	The calculated spectra retain the characteristic \ce{N-N} $\pi^*$ resonance, while
	its intensity and line shape vary substantially with the surrounding
	Ga-defect environment. This variation is consistent with the
	temperature-dependent experimental response, in which the molecular \ce{N_{2}}
	signature is weaker in the as-implanted state and becomes progressively
	more pronounced upon annealing \cite{7}. The present
	calculations suggest that, in the highly damaged as-implanted material,
	nitrogen may form \ce{N-N} bonded complexes involving both interstitial and
	substitutional N atoms, stabilized by nearby Ga vacancies and
	Frenkel-like rearrangements. Such configurations are consistent with the
	high concentration of intrinsic defects generated during ion
	implantation. Ga Frenkel pairs are predicted to form under displacement
	damage in \ce{\beta-Ga2O3} and have been implicated in the irradiation-induced
	\ce{\beta}-to-\ce{\gamma} structural transformation \cite{27, 28}. Upon
	annealing, increasing defect mobility promotes vacancy-interstitial
	migration and recombination, progressively reducing the population of
	implantation-induced defects \cite{28,31,32}. The resulting
	structural recovery, together with nitrogen redistribution, may reduce
	the relative stability or population of Frenkel-associated \ce{N-N} complexes
	and favor the formation of more strongly molecularized nitrogen
	configurations. This interpretation is consistent with the
	experimentally established evolution toward molecular \ce{N_{2}} upon
	annealing \cite{7}. The present results therefore provide an
	additional atomistic perspective on this process, suggesting that
	substitutional-interstitial N pairs associated with Ga defects may
	represent intermediate configurations in the thermal evolution of
	implanted nitrogen. The proposed involvement of these specific complexes
	remains a mechanistic hypothesis rather than a unique assignment of the
	experimental spectra.

\subsection{Localized Gap States and Spin Polarization}
\label{section3.3}

\begin{figure}[!htpb]
		\centering
		\begin{subfigure}{0.47\textwidth}
			\centering
			\includegraphics[width=\textwidth]{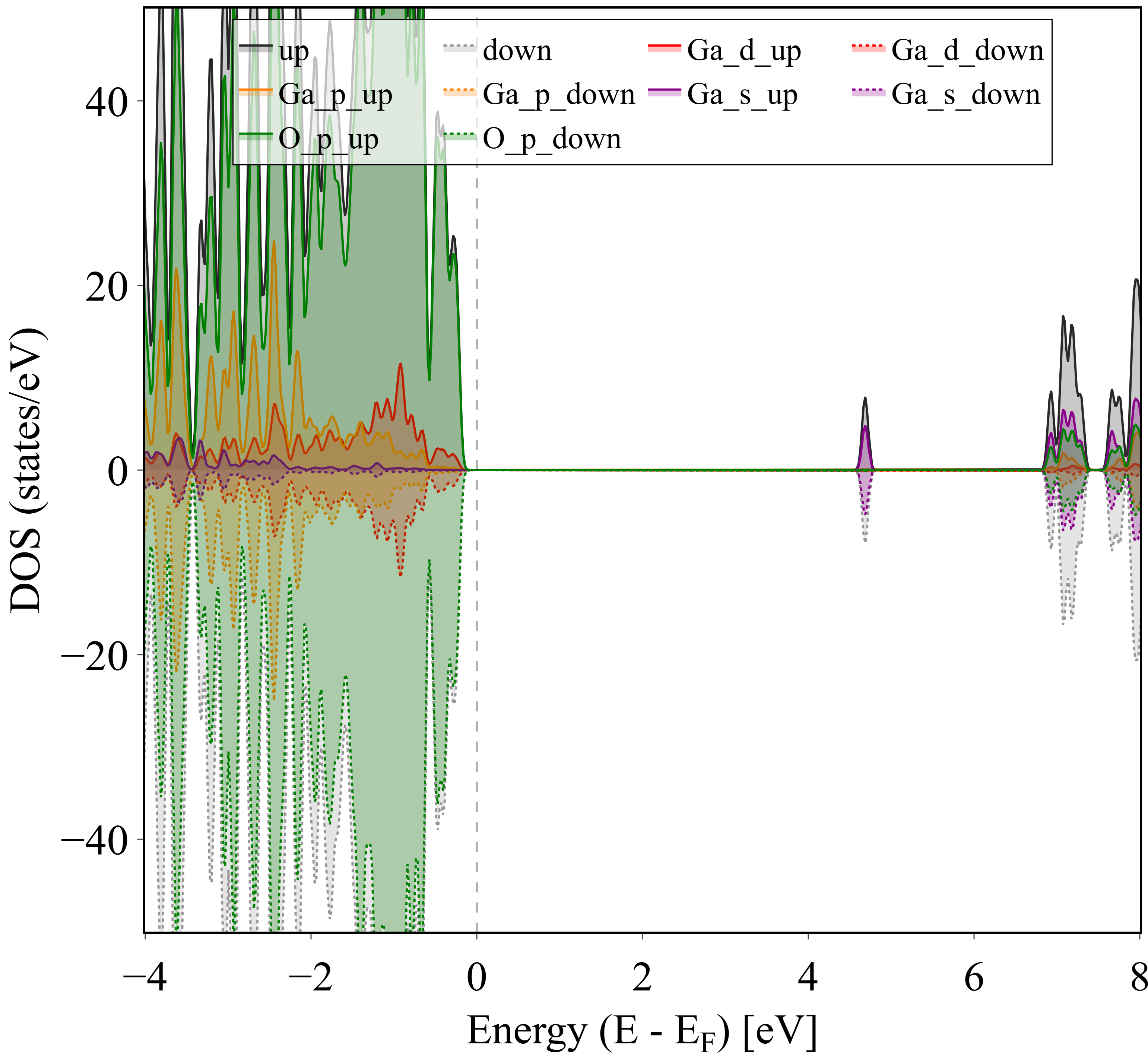}
			\caption{pristine}
			\label{figure5a}
		\end{subfigure}
		\begin{subfigure}{0.47\textwidth}
			\centering
			\includegraphics[width=\textwidth]{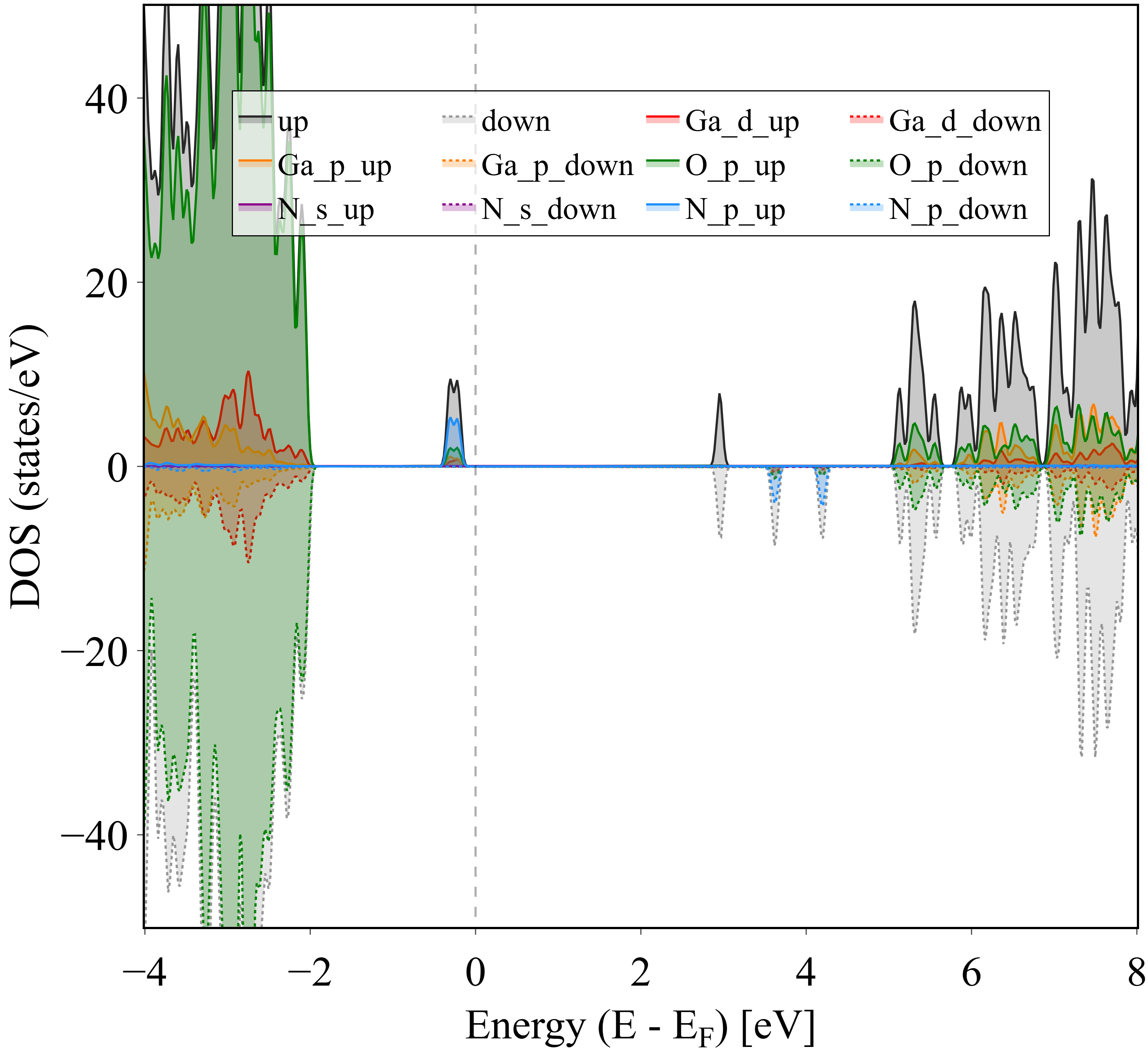}
			\caption{\ce{N_{i9}-N_{OI}}}
			\label{figure5b}
		\end{subfigure}
		\caption{
Total and projected densities of states (TDOS/PDOS) for 
(a) pristine \ce{\beta-Ga2O3} and 
(b) \ce{N_{i9}-N_{OI}}. 
The energy zero is the Fermi energy. 
Nitrogen-derived states occur within the band gap.
		} 
		\label{figure5}
	\end{figure}

	To clarify the nature and spatial character of the defect-induced
	states, the electronic structure of the considered nitrogen-related
	defect complexes in \ce{\beta-Ga2O3} was further
	analyzed using the density of states (DOS) and spin density
	distributions. The DOS results show that the introduction of nitrogen
	dopants and intrinsic vacancies leads to the emergence of electronic
	states within the band gap, confirming the formation of deep-level
	defects. The valence band remains predominantly composed of O-$2p$ states,
	while the conduction band is mainly derived from Ga-$4s$ states,
	consistent with pristine \ce{\beta-Ga2O3} (Figure~\ref{figure5a}). 
	Upon nitrogen incorporation in the form of the
	\ce{N_{i9}-N_{OI}} defect complex, additional
	localized states are introduced within the band gap, as clearly observed
	in Figure~\ref{figure5b}. 
	These states are dominated by N-$2p$ contributions and are
	separated from the band edges, indicating localized trap-like behavior
	rather than that of a shallow band-edge state.

	\begin{figure}[!htpb]
		\centering
		\begin{subfigure}{0.47\textwidth}
			\centering
			\includegraphics[width=\textwidth]{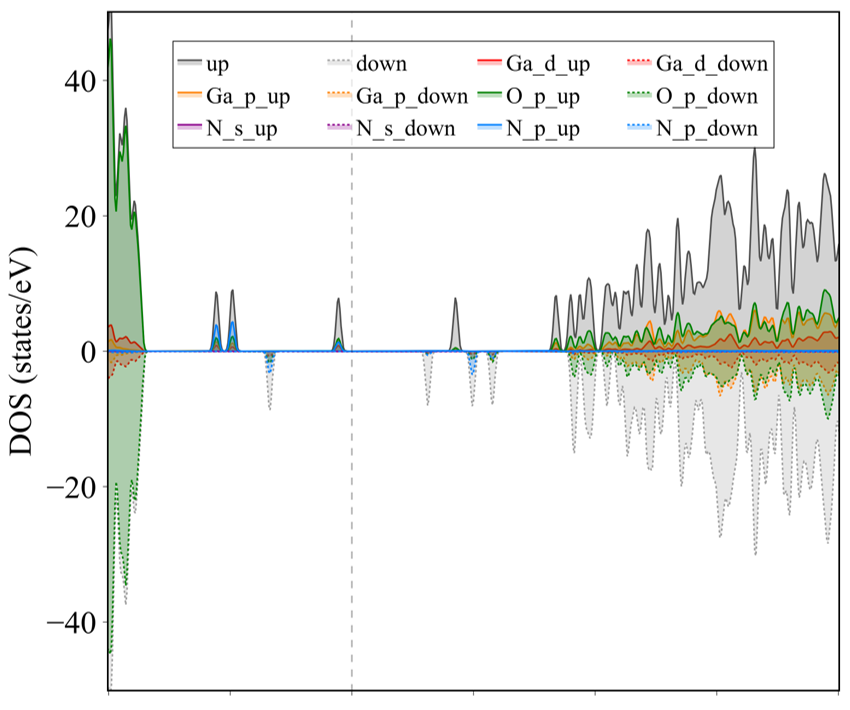}
			\caption{\ce{N_{i9}-N_{OI}-V_{OI}}}
			\label{figure6a}
		\end{subfigure}
		\begin{subfigure}{0.47\textwidth}
			\centering
			\includegraphics[width=\textwidth]{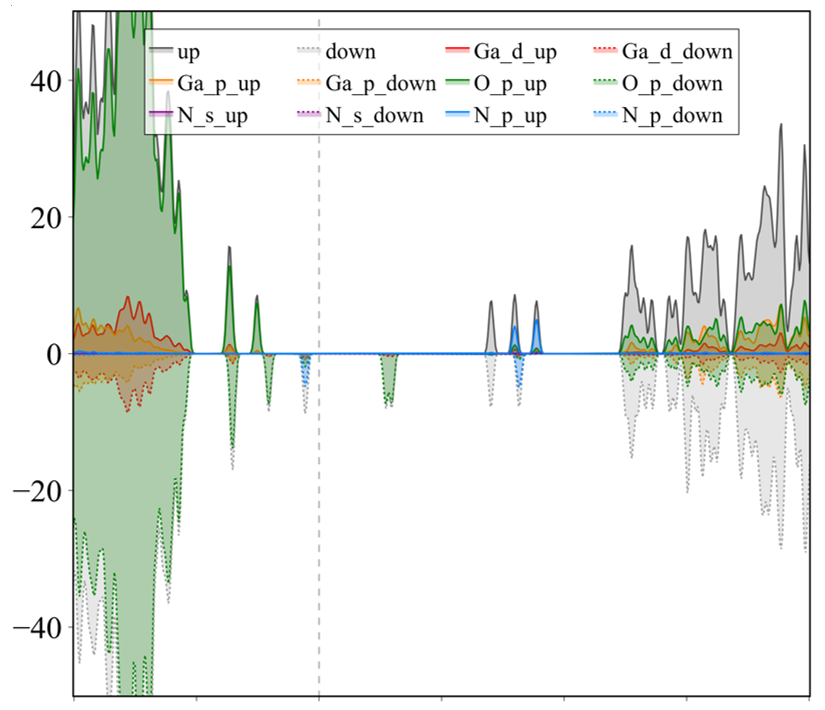}
			\caption{\ce{N_{i9}-N_{OI}-V_{GaI}}}
			\label{figure6b}
		\end{subfigure}\\
		\begin{subfigure}{0.47\textwidth}
			\centering
			\includegraphics[width=\textwidth]{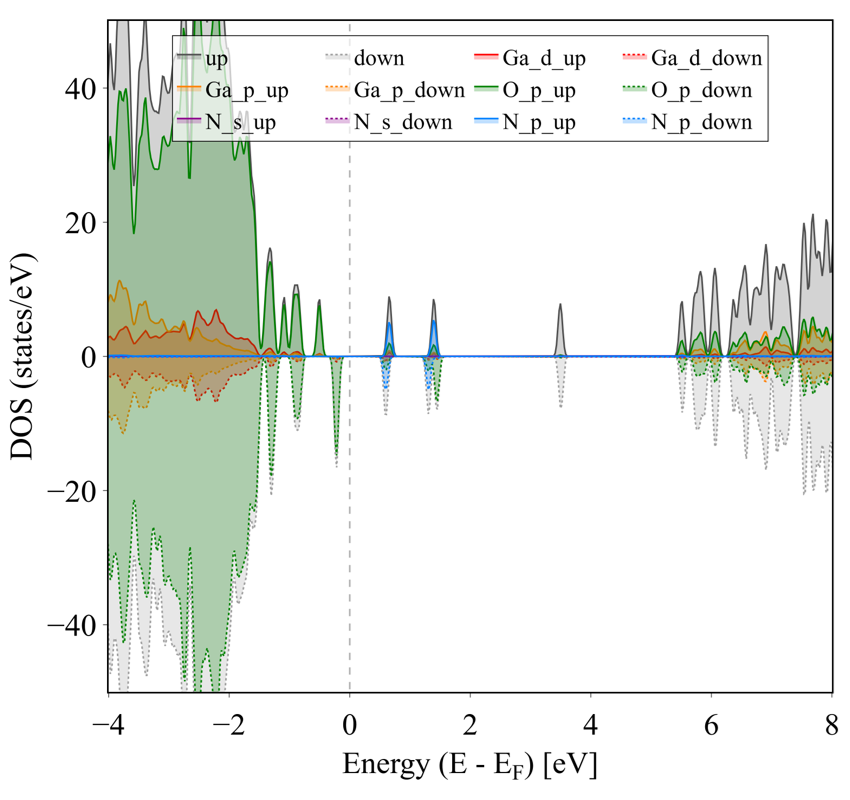}
			\caption{\ce{N_{i9}-N_{OI}-V_{GaII}}}
			\label{figure6c}
		\end{subfigure}
		\begin{subfigure}{0.47\textwidth}
			\centering
			\includegraphics[width=\textwidth]{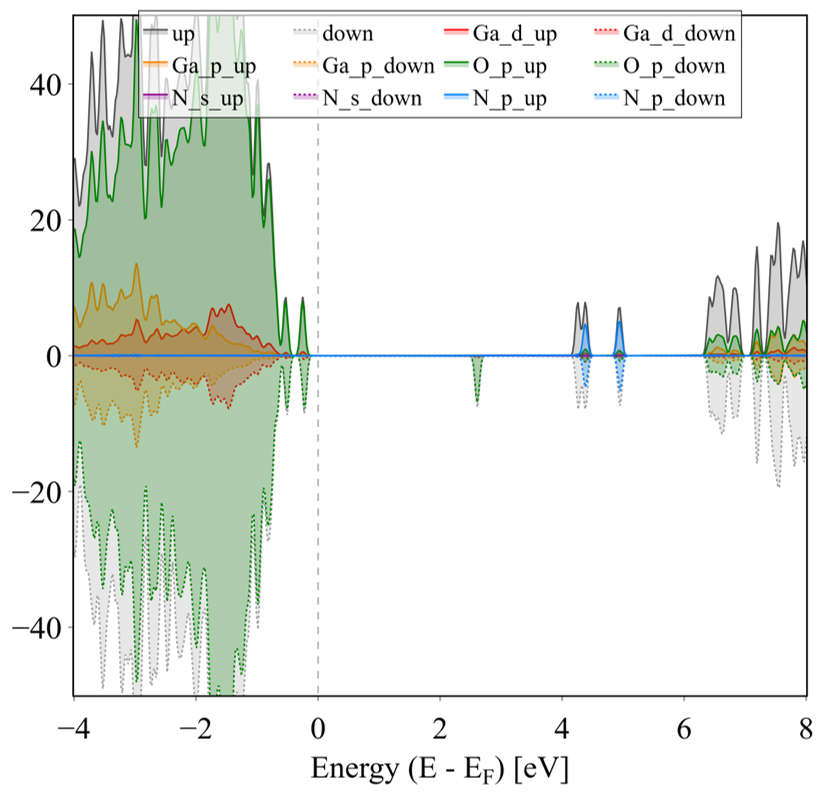}
			\caption{\ce{N_{i9}-N_{OI}-V_{GaII}-{FP}_{GaI}}}
			\label{figure6d}
		\end{subfigure}\\
		\begin{subfigure}{0.47\textwidth}
			\centering
			\includegraphics[width=\textwidth]{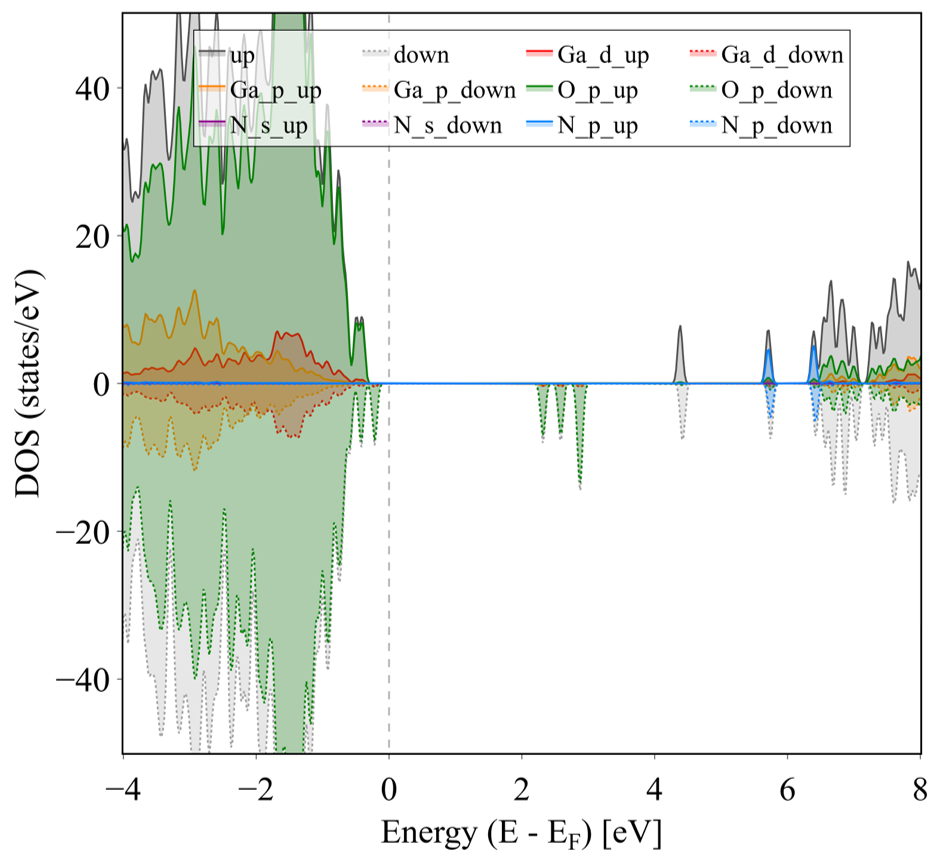}
			\caption{\ce{N_{i9}-N_{OI}-V_{GaI}-V_{GaII}}}
			\label{figure6e}
		\end{subfigure}
		\caption{
Total and projected densities of states (TDOS/PDOS) 
for \ce{\beta-Ga2O3} containing 
(a) \ce{N_{i9}-N_{OI}-V_{OI}}, 
(b) \ce{N_{i9}-N_{OI}-V_{GaI}}, 
(c) \ce{N_{i9}-N_{OI}-V_{GaII}}, 
(d) \ce{N_{i9}-N_{OI}-V_{GaII}-{FP}_{GaI}}, and 
(e) \ce{N_{i9}-N_{OI}-V_{GaI}-V_{GaII}}.
The energy zero is the Fermi energy.
		} 
		\label{figure6}
	\end{figure}
	
	The O-vacancy-assisted complex shifts the calculated Fermi level away
	from the VBM, consistent with the donor-like character of
	\ce{V_{OI}} (Figure~\ref{figure6a}). 
	Nevertheless, localized
	states dominated by N-$2p$ contributions remain deep within the band gap.
	These states may act as carrier traps and potential recombination
	centers, which could limit carrier transport and reduce electrical
	conductivity.

	In contrast, Ga-vacancy-assisted complexes shift the calculated Fermi
	level toward the VBM and introduce unoccupied localized states dominated
	by O-$2p$ contributions (Figure~\ref{figure6b}-\ref{figure6e}). 
	Although this behavior is acceptor-like, the
	states remain deep within the band gap and spatially localized. They can
	accept electrons but do not produce the delocalized valence-band holes
	required for shallow-acceptor behavior and effective $p$-type transport.
	
\begin{figure}[!htbp]
		\centering
		\begin{subfigure}{0.3\textwidth}
			\centering
			\includegraphics[width=\textwidth]{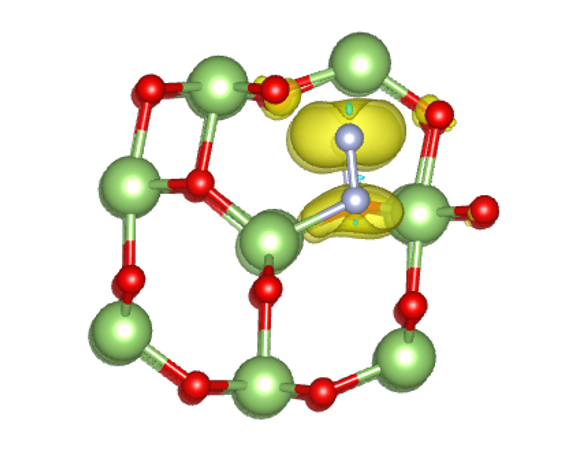}
			\caption{\ce{N_{i9}-N_{OI}}}
			\label{figure7a}
		\end{subfigure}
		\begin{subfigure}{0.3\textwidth}
			\centering
			\includegraphics[width=\textwidth]{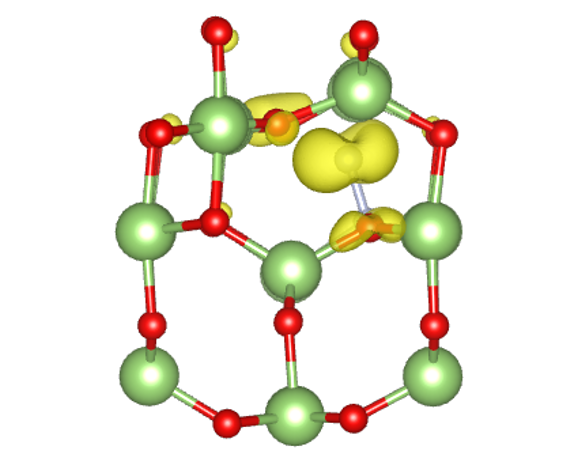}
			\caption{\ce{N_{i9}-N_{OI}-V_{OI}}}
			\label{figure7b}
		\end{subfigure}
		\begin{subfigure}{0.3\textwidth}
			\centering
			\includegraphics[width=\textwidth]{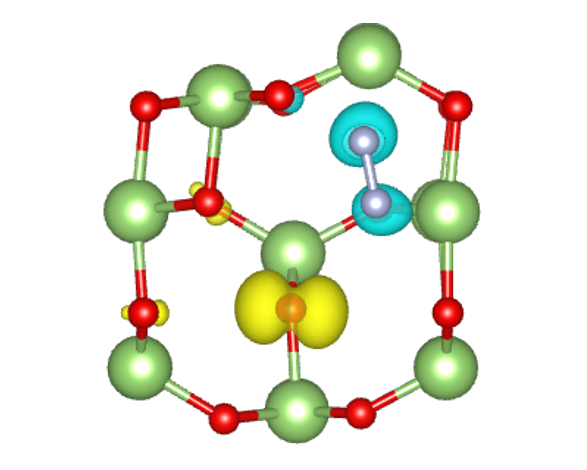}
			\caption{\ce{N_{i9}-N_{OI}-V_{GaI}}}
			\label{figure7c}
		\end{subfigure} \\
		\begin{subfigure}{0.3\textwidth}
			\centering
			\includegraphics[width=\textwidth]{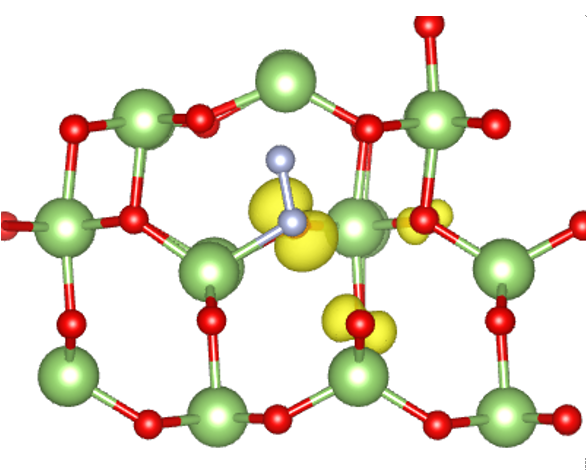}
			\caption{\ce{N_{i9}-N_{OI}-V_{GaII}}}
			\label{figure7d}
		\end{subfigure}
		\begin{subfigure}{0.3\textwidth}
			\centering
			\includegraphics[width=\textwidth]{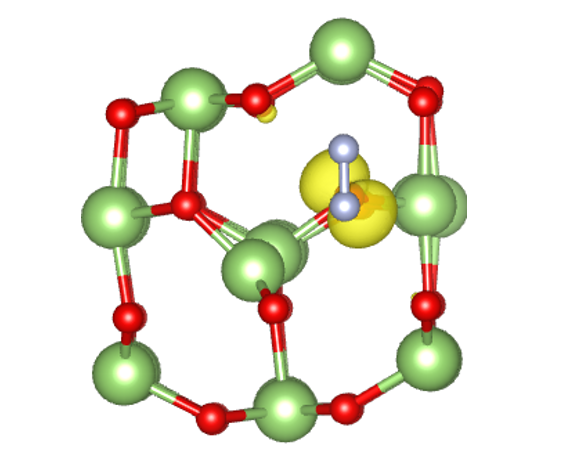}
			\caption{\ce{N_{i9}-N_{OI}-V_{GaII}-{FP}_{GaI}}}
			\label{figure7e}
		\end{subfigure}
		\begin{subfigure}{0.3\textwidth}
			\centering
			\includegraphics[width=\textwidth]{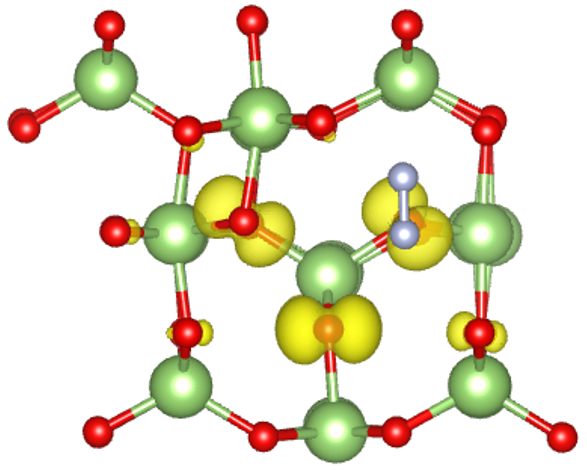}
			\caption{\ce{N_{i9}-N_{OI}-V_{GaI}-V_{GaII}}}
			\label{figure7f}
		\end{subfigure}
		\caption{
Spin-density isosurfaces for 
(a) \ce{N_{i9}-N_{OI}}, 
(b) \ce{N_{i9}-N_{OI}-V_{OI}}, 
(c) \ce{N_{i9}-N_{OI}-V_{GaI}}, 
(d) \ce{N_{i9}-N_{OI}-V_{GaII}}, 
(e) \ce{N_{i9}-N_{OI}-V_{GaII}-{FP}_{GaI}}, and 
(f) \ce{N_{i9}-N_{OI}-V_{GaI}-V_{GaII}}.
Blue and yellow isosurfaces denote opposite spin polarization. 
For each panel, the isovalue is \qty{1}{\percent} of the maximum absolute spin density
		} 
		\label{figure7}
	\end{figure}

	The spin density analysis shown in Figure~\ref{figure7} further reveals a strong
	dependence of electronic localization on the local defect environment.
	For isolated nitrogen interstitial-substitution complexes, the spin
	density is strongly localized around the N sites, indicating a
	well-defined defect-centered magnetic moment. However, in Ga
	vacancy-assisted configurations, the spin density redistributes from the
	nitrogen sites toward neighboring oxygen atoms, reflecting enhanced
	defect-host hybridization and lattice-mediated charge redistribution. In
	the more complex configurations, the spin polarization extends over
	several neighboring atoms but remains confined to the defect region.
	Together, the DOS, projected DOS (PDOS), 
	and spin-density results show that the defect
	states lie deep within the band gap and remain spatially localized,
	supporting their assignment as deep trapping states that can limit
	carrier transport in \ce{\beta-Ga2O3}.

\section{Conclusions}
\label{section4}
	
Hybrid-functional calculations reveal a strong tendency for interstitial
and substitutional N atoms to co-localize in \ce{\beta-Ga2O3}, forming the
energetically favorable \ce{N_{i9}-N_{OI}} pair.
Nearby native vacancies strongly modify this interaction in a
vacancy-specific manner. While \ce{V_{OI}} increases the \ce{N-N}
separation, selected Ga-vacancy environments reduce local structural
constraints and shorten the \ce{N-N} distance to \qtyrange{1.09}{1.19}{\angstrom}. 
The corresponding ELF distributions indicate substantial \ce{N-N} covalent
bonding. Positive binding energies show that the complexes are bound
relative to their isolated constituents, whereas their often high
formation energies suggest that the more complex structures are most
relevant as metastable products of implantation rather than abundant
equilibrium defects. Although selected Ga-vacancy configurations shift
the Fermi level toward the valence-band maximum, none of the
configurations examined produces a shallow acceptor. Instead, localized
N $2p$- and O $2p$-derived states persist within the band gap, with spin
polarization concentrated around the defect complexes and neighboring O
atoms. Calculated N~$K$-edge XANES spectra for two selected
Ga-defect-assisted complexes demonstrate that the spectral
characteristics of \ce{N-N} bonding are sensitive to the surrounding defect
environment. Together with the experimentally established formation of
molecular \ce{N_{2}} upon annealing, these results suggest that
substitutional-interstitial N pairs associated with Ga vacancies and
Frenkel-like rearrangements may represent intermediate configurations in
the structural evolution of implanted nitrogen. These results link
vacancy-assisted \ce{N-N} bonding and defect-complex formation to deep
carrier trapping, providing an atomistic explanation consistent with the
semi-insulating, current-blocking behavior of N-implanted \ce{\beta-Ga2O3}.

\newpage

\section*{Author Contributions} 
	Conceptualization, A.S. and I.N.D.; 
	data curation, A.S.; 
	formal analysis, A.S.; 
	funding acquisition, I.N.D.; 
	methodology, A.S., Y.M., Y.S., M.C., and I.N.D.; 
	project administration, I.N.D.; 
	resources, A.S., Y.S., and I.N.D.; 
	software, A.S. and Y.S.; 
	supervision, I.N.D.;
	validation, A.S., Y.M., Y.S., M.C., and I.N.D.; 
	visualization, A.S.;
	writing---original draft, A.S.; 
	writing---review and editing, A.S., Y.M., M.C., and I.N.D. 
	All authors have approved the submitted version of the manuscript.
	
\section*{Funding:} 
	This research was funded by National Science Centre (NCN) in Poland, grant UMO-2020/39/B/ST5/03580.
	
\section*{Data Availability Statement:} 
	The data supporting the findings of this study will be openly available
	in RepOD-Repository for Open Data at
	\url{https://doi.org/10.18150/CVDW9G}.

\section*{Acknowledgments} 
	This work was supported by the National Science Centre, Poland, under
	grant UMO-2020/39/B/ST5/03580. Computational resources were provided by
	the Interdisciplinary Centre for Mathematical and Computational
	Modelling, University of Warsaw, under allocation G35-57. This research
	was also partially supported by the Warsaw University of Technology
	Excellence Initiative: Research University (IDUB) program. A.S also
	acknowledges support from the ``MagTop'' project
	(FENG.02.01-IP.05-0028/23) carried out within the ``International
	Research Agendas'' programme of the Foundation for Polish Science,
	co-financed by the European Union under the European Funds for Smart
	Economy 2021-2027 (FENG). During the preparation of this manuscript, the
	authors used Gemini 3.1 Pro and DeepL for assistance in Language \&
	Writing (paraphrasing and rewording, improving grammar and language,
	adjusting tone and style) and Review \& Compliance (checking manuscript
	quality). The authors have reviewed and edited the output and take full
	responsibility for the content of this publication.
	
\section*{Conflicts of Interest:} 
	The authors declare no competing financial interests.
	
\section*{Abbreviations}
	
	The following abbreviations are used in this manuscript:
	
	\begin{tabular}{ll}
		DFT   & Density Functional Theory \\
		DOS   & Density of states \\
		ELF   & Electron localization function \\
		GGA   & Generalized gradient approximation \\
		HSE06 & Heyd--Scuseria--Ernzerhof hybrid functional \\
		PAW   & Projector augmented wave method \\
		PBE   & Perdew-Burke-Ernzerhof exchange-correlation functional \\
		PDOS  & Projected density of states \\
		RT   & Room temperature \\
		TDOS  & Total density of states \\
		TFY  & Total fluorescence yield \\
		UWBG  & Ultra-wide band gap \\
		VASP  & The Vienna Ab initio Simulation Package \\
		VBM   & Valence band maximum \\
		XANES & X-ray absorption near-edge structure
	\end{tabular}
	
\bibliographystyle{unsrtnat}

\bibliography{references.bib}

\end{document}